\documentclass[twocolumn, times]{aastex62}
\usepackage{aas_macros, graphicx}

\def\lesssim{\mathrel{\hbox{\rlap{\hbox{\lower4pt\hbox{$\sim$}}}\hbox{$<$}}}}
\def\gtrsim{\mathrel{\hbox{\rlap{\hbox{\lower4pt\hbox{$\sim$}}}\hbox{$>$}}}}
\received{--}
\revised{--}
\accepted{--}
\submitjournal{ApJ}

\shorttitle{SN 2011dh at late-times}
\shortauthors{Maund}

\begin{document}

\title{The origin of the late-time luminosity of supernova 2011dh}

\correspondingauthor{Justyn R. Maund}
\email{j.maund@sheffield.ac.uk}

\author[0000-0002-0786-7307]{Justyn R. Maund}
\affil{Department of Physics and Astronomy\\
University of Sheffield\\
Hicks Building, Hounsfield Road\\
Sheffield S3 7RH, U.K.}


\begin{abstract}

Due to the small amount of hydrogen (${\leq 0.1M_{\odot}}$) remaining on the surface of their progenitors, Type IIb supernovae are sensitive probes of the mass loss processes of massive stars towards the ends of  their lives, including the role of binarity.  We report late-time Hubble Space Telescope observations of SN 2011dh in M51, and a brief period of re-brightening and plateau in the photometric light curve, from $1.8$ to $6.2$ years after the explosion.  These observations exclude the role of circumstellar interaction, however a slow rotating magnetar, a significant quantity of radioactive elements or a light echo could be responsible for the late-time luminosity observed at $t > 1000\mathrm{d}$.  If the late-time light curve is powered by the decay of radioactive elements, SN~2011dh is required to have produced $\sim 2.6 \times 10^{-3}\,M_{\odot}$ of $\mathrm{^{44}Ti}$, which is significantly in excess of the amount inferred from earlier nebular spectra of SN 2011dh itself or measured in the Cas A SN remnant.  The evolution of the brightness and the colour of the late-time light curve also supports the role of a light echo originating from dust with a preferred geometry of a disk of extent $\sim 1.8$ to $\sim 2.7\,\mathrm{pc}$ from the SN, consistent with a wind-blown bubble.  Accounting for the long term photometric evolution due to a light echo, the flux contribution from a surviving binary companion at ultraviolet wavelengths can be isolated and corresponds to a star of $\sim  9 - 10M_{\odot}$.

\end{abstract}

\keywords{supernovae: individual(2011dh) -- stars: massive}

%
%
\section{Introduction}
\label{sec:intro}
Type IIb supernovae (SNe) are one of the most interesting classes of exploding massive stars that are observable in the near universe.  Exhibiting spectral features due to the presence of hydrogen in their early time spectra, these features quickly grow weaker until they completely disappear and the SN transitions to a hydrogen-deficient Type Ib SN \citep{fili97}.  The evolution of such SNe is indicative of the progenitor stars having been stripped of all but a minute amount of hydrogen ($<0.1M_{\odot}$) and, as such, these SNe and their progenitors provide exquisite probes of the mass loss processes that massive stars may undergo \citep{2014ARA&A..52..487S}.  The progenitor of the class prototype, SN~1993J, was observed to be a K-supergiant, whose appearance did not fit the canonical predictions of single-star stellar evolution models \citep{alder93j}.  It was not until 2004, with the detection of the binary companion in the fading remnant, that the origin of the unusual properties of this SN and its progenitor through binary interaction were confirmed \citep{maund93j, 2014ApJ...790...17F}.  More recently, a hot companion star was recovered at the position of the Type IIb SN 2001ig \citep{2018ApJ...856...83R}.

\begin{figure}
\begin{center}
\includegraphics[width=0.8\linewidth]{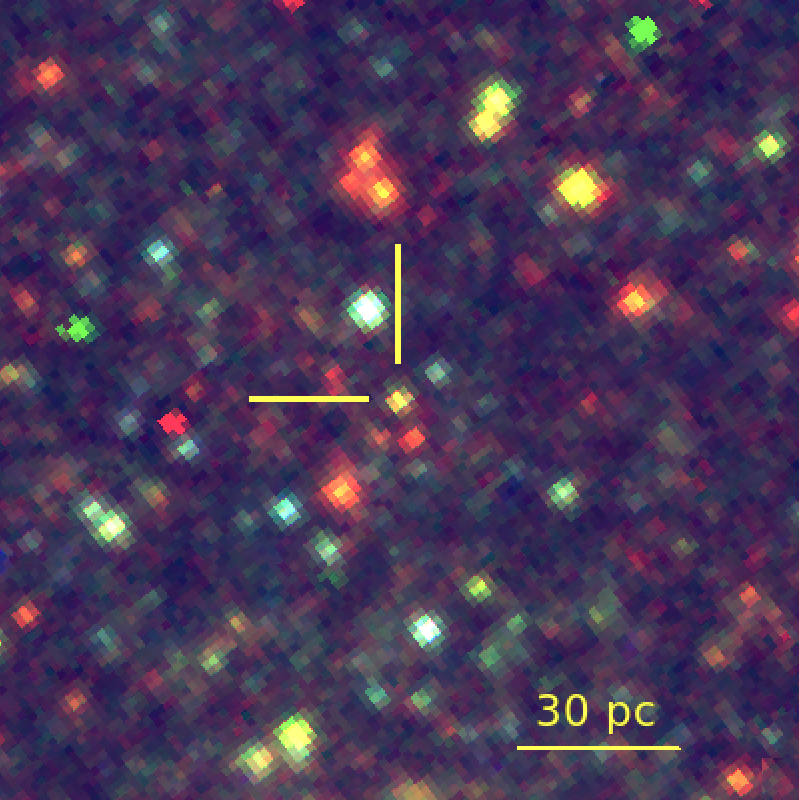}
\end{center}
	\caption{\small{Late-time HST WFC3 UVIS observation of SN 2011dh in M51 at 8 Aug 2017 ($2261\mathrm{d}$ post-explosion).  The image has dimensions $150 \times 150\,\mathrm{pc}$, and is oriented with North up and East left (with the SN indicated by the cross-hairs).  The three-colour image is composed of the $F336W$, $F555W$ and $F814W$ broad-band observations.}}
\label{fig:picture}
\end{figure}

In the last decade, SN 2011dh has been a landmark Type IIb SN, having been discovered within $\sim 3\,{\mathrm{days}}$ of explosion in the nearby galaxy M51 \citep{CBET2736,2011ApJ...742L..18A} (see Figure \ref{fig:picture}). Its proximity \citep[distance modulus $\mu = 29.46 \pm 0.28\,\mathrm{mags}$;][]{2014A&A...562A..17E} meant that SN~2011dh was subject to massive multi-wavelength followup campaigns \citep{2014A&A...562A..17E,2012ApJ...751..125B,2012ApJ...752...78S,2013MNRAS.436.1258H,2013ApJ...778L..19H} and a progenitor was observed in fortuitous pre-explosion Hubble Space Telescope (HST) observations \citep{2011ApJ...739L..37M, 2011ApJ...741L..28V,  2012ApJ...747...23S}.  The progenitor candidate was identified as a Yellow Supergiant (YSG), arising from a $13M_{\odot}$ star; however, the early light curve evolution of the SN, with the fast decaying cooling phase, suggested that the star that exploded was a fainter, more compact companion to the YSG \citep{2011ApJ...742L..18A,2012ApJ...752...78S}.  Only later observations, showing the disappearance of the YSG, confirmed the detected star did explode \citep{2014A&A...562A..17E,2013ApJ...772L..32V}.  

Single and binary progenitor scenarios for SN 2011dh have been debated \citep{2012A&A...538L...8G,2013ApJ...762...74B}, with significant implications for how massive stars lose mass \citep[in particular the contrast between eruptive episodes of intense mass loss and steady mass loss through sustained winds;][]{2014ARA&A..52..487S} and for the origins of SNe deficient in hydrogen and helium.  A search for the companion in the fading remnant of SN~2011dh in 2014 was inconclusive \citep{2014ApJ...793L..22F,2015MNRAS.454.2580M}, primarily because the SN itself had not faded away.  Here, we present more recent HST observations of the site of SN~2011dh to identify the cause of the continuing brightness of the SN up to $2300\mathrm{d}$ after explosion.

%
%

\section{Observations and Data Reduction}
\label{sec:obs}

To probe the late-time evolution of the brightness of SN~2011dh, we consider a series of ultraviolet (UV) and optical HST observations spanning $640$ to $2299\mathrm{d}$ post-explosion, at 38 separate epochs, assuming an explosion epoch of $t_{\mathrm{exp}} = \mathrm{MJD}55712.5$ or 2011 May 31.5 UT \citep{2014A&A...562A..17E}.  A log of these late-time HST observations of SN~2011dh is presented in Table \ref{tab:obs:log}.  The observations were conducted using the Advanced Camera for Surveys (ACS) Wide Field Channel (WFC) and the Wide Field Camera 3 (WFC3) Ultraviolet-Visible (UVIS) channel.  All observations were retrieved from the Milkuski Archive for Space Telescopes\footnote{https://archive.stsci.edu/hst/}.  All the images were aligned to a common astrometric reference frame, defined by the ACS WFC observations from 2016 May 5, using the {\sc PyRAF} package {\tt tweakreg}\footnote{PyRAF is a product of the Space Telescope Science Institute, which is operated by AURA for NASA} task, with the position of SN~2011dh determined with respect to the location identified by \citet{2011ApJ...739L..37M}.  

Photometry of the HST observations was conducted using the DOLPHOT package \citep{dolphhstphot, 2016ascl.soft08013D}\footnote{http://americano.dolphinsim.com/dolphot/}.  We report photometry in the Vega magnitude system.  DOLPHOT photometry, especially in crowded fields, can be sensitive to the treatment of the sky background. Photometry was conducted, therefore, using two settings for accounting for the sky background to assess any systematic errors due to handling of the background. For the 2016 and 2017 ACS WFC $F606W$ and $F814W$ observations, we find differences between the ``FITSKY = 2" and ``FITSKY = 3" photometry modes to be $< 0.06\,\mathrm{mags}$, and report the latter.  For the later WFC3 UVIS observations, we find that the differences between the ``FITSKY = 2" and ``FITSKY = 3" photometry are much less than the photometric uncertainty, even in the case of the narrow-band observations for which the signal-to-noise levels are significantly lower.

\startlongtable
\begin{deluxetable*}{llcccl}
    \tablecaption{Late-time HST observations of the site of SN 2011dh \label{tab:obs:log}}  
\tablehead{
\colhead{Dataset} & \colhead{Date (UT)} & \colhead{Instrument} & \colhead{Exp time(s)} & \colhead{Filter} & \colhead{Program}}
\startdata
IBYB06010     &2013-03-02              &WFC3 UVIS1   &  680        & F814W   & 13029\tablenotemark{a}  \\
IBYB06020     &2013-03-02              &WFC3 UVIS1   &  510        & F555W   & 13029     \\
\\
IC7S01010     &2014-08-07              &WFC3 UVIS2   & 3772        & F225W   & 13433\tablenotemark{b}  \\  
IC7S01020     &2014-08-07              &WFC3 UVIS2   & 1784        & F336W   & 13433    \\
JC7S02010     &2014-08-10              &ACS  WFC1    & 1072        & F435W   & 13433    \\
JC7S02020     &2014-08-10              &ACS  WFC1    & 1232        & F555W   & 13433    \\
JC7S02030     &2014-08-10              &ACS  WFC1    & 2176        & F814W   & 13433    \\
\\
ICUQ31010     &2016-01-01              &WFC3 UVIS1   & 780         & F336W   & 14149\tablenotemark{a}\\     
ICUQ31020     &2016-01-01              &WFC3 UVIS1   & 710         & F555W   & 14149    \\
\\
JD8F01010     &2016-10-05              &ACS          & 2200        & F606W   &14704\tablenotemark{c}  \\    
JD8F01020     &2016-10-05              &ACS          & 2200        & F814W   &14704      \\
JD8F02010     &2016-10-14              &ACS          & 2200        & F606W   &14704      \\
JD8F02020     &2016-10-14              &ACS          & 2200        & F814W   &14704      \\
JD8F03010     &2016-11-07              &ACS          & 2200        & F606W   &14704      \\
JD8F03020     &2016-11-07              &ACS          & 2200        & F814W   &14704      \\
JD8F04010     &2016-11-11              &ACS          & 2200        & F606W   &14704      \\
JD8F04020     &2016-11-11              &ACS          & 2200        & F814W   &14704      \\
JD8F05010     &2016-11-27              &ACS          & 2200        & F606W   &14704      \\
JD8F05020     &2016-11-27              &ACS          & 2200        & F814W   &14704      \\
JD8F06010     &2016-12-07              &ACS          & 2200        & F606W   &14704      \\
JD8F06020     &2016-12-07              &ACS          & 2200        & F814W   &14704      \\
JD8F07010     &2016-12-31              &ACS          & 2200        & F606W   &14704      \\
JD8F07020     &2016-12-31              &ACS          & 2200        & F814W   &14704      \\
JD8F08010     &2017-01-07              &ACS          & 2200        & F606W   &14704      \\
JD8F08020     &2017-01-07              &ACS          & 2200        & F814W   &14704      \\
JD8F10010     &2017-01-20              &ACS          & 2200        & F606W   &14704      \\
JD8F10020     &2017-01-20              &ACS          & 2200        & F814W   &14704      \\
JD8F11010     &2017-01-29              &ACS          & 2200        & F606W   &14704      \\
JD8F11020     &2017-01-29              &ACS          & 2200        & F814W   &14704      \\
JD8F35010     &2017-02-17              &ACS          & 2150        & F606W   &14704      \\
JD8F35020     &2017-02-17              &ACS          & 2200        & F814W   &14704      \\
JD8F12010     &2017-03-04              &ACS          & 2200        & F606W   &14704      \\
JD8F12020     &2017-03-04              &ACS          & 2200        & F814W   &14704      \\
JD8F13010     &2017-03-11              &ACS          & 2200        & F606W   &14704      \\
JD8F13020     &2017-03-11              &ACS          & 2200        & F814W   &14704      \\
JD8F14010     &2017-03-21              &ACS          & 2200        & F606W   &14704      \\
JD8F14020     &2017-03-21              &ACS          & 2200        & F814W   &14704      \\
JD8F15010     &2017-03-25              &ACS          & 2200        & F606W   &14704      \\
JD8F15020     &2017-03-25              &ACS          & 2200        & F814W   &14704      \\
JD8F16010     &2017-04-15              &ACS          & 2200        & F606W   &14704      \\
JD8F16020     &2017-04-15              &ACS          & 2200        & F814W   &14704      \\
JD8F17010     &2017-04-20              &ACS          & 2200        & F606W   &14704      \\
JD8F17020     &2017-04-20              &ACS          & 2200        & F814W   &14704      \\
JD8F18010     &2017-04-28              &ACS          & 2200        & F606W   &14704      \\
JD8F18020     &2017-04-28              &ACS          & 2200        & F814W   &14704      \\
JD8F19010     &2017-05-09              &ACS          & 2200        & F606W   &14704      \\
JD8F19020     &2017-05-09              &ACS          & 2200        & F814W   &14704      \\
JD8F20010     &2017-05-28              &ACS          & 2200        & F606W   &14704      \\
JD8F20020     &2017-05-28              &ACS          & 2200        & F814W   &14704      \\
JD8F21010     &2017-06-07              &ACS          & 2200        & F606W   &14704      \\
JD8F21020     &2017-06-07              &ACS          & 2200        & F814W   &14704      \\
JD8F22010     &2017-06-11              &ACS          & 2084        & F606W   &14704      \\
JD8F22020     &2017-06-11              &ACS          & 2200        & F814W   &14704      \\
JD8F23010	  &2017-06-21              &ACS	         & 2200	       & F606W	 &14704		\\	    
JD8F23020	  &2017-06-21              &ACS	         & 2200	       & F814W	 &14704		\\	    
JD8F24010	  &2017-06-27              &ACS	         & 2200	       & F606W	 &14704		\\	    
JD8F24020	  &2017-06-27              &ACS	         & 2200	       & F814W	 &14704		\\	    
JD8F25010	  &2017-07-02              &ACS	         & 2200	       & F606W	 &14704		\\	    
JD8F25020	  &2017-07-02              &ACS	         & 2200	       & F814W	 &14704		\\	    
JD8F26010	  &2017-07-12              &ACS	         & 2200	       & F606W	 &14704		\\	    
JD8F26020	  &2017-07-12	           &ACS	         & 2200	       & F814W	 &14704		\\	   
JD8F27010	  &2017-07-19              &ACS	         & 2200	       & F606W	 &14704		\\	  
JD8F27020	  &2017-07-19              &ACS	         & 2200	       & F814W	 &14704		\\	  
JD8F28010     &2017-07-30              &ACS          & 2200        & F606W   &14704     \\
JD8F28020     &2017-07-30              &ACS          & 2200        & F814W   &14704     \\
JD8F29010     &2017-08-06              &ACS          & 2200        & F606W   &14704     \\
JD8F29020     &2017-08-06              &ACS          & 2200        & F814W   &14704     \\
JD8F30010	  &2017-08-13              &ACS	         & 2200	       & F606W	 &14704		\\	   
JD8F30020	  &2017-08-13              &ACS	         & 2200	       & F814W	 &14704		\\	   
JD8F31010	  &2017-08-20              &ACS		     & 2200	       & F606W	 &14704		\\	   
JD8F31020	  &2017-08-20              &ACS		     & 2200	       & F814W	 &14704		\\	   
JD8F32010	  &2017-08-28              &ACS		     & 2200	       & F606W	 &14704		\\	   
JD8F32020	  &2017-08-28              &ACS		     & 2200	       & F814W	 &14704		\\	   
JD8F33010	  &2017-09-03              &ACS		     & 2200	       & F606W	 &14704		\\	   
JD8F33020	  &2017-09-03              &ACS		     & 2200	       & F814W	 &14704		\\	   
JD8F34010	  &2017-09-15              &ACS		     & 2200	       & F606W	 &14704		\\	   
JD8F34020	  &2017-09-15              &ACS	         & 2200	       & F814W	 &14704		\\  
\\
ID5P01010     &2017-08-08              &WFC3 UVIS2   &2020         & F336W   &14763\tablenotemark{b}      \\
ID5P01020     &2017-08-08              &WFC3 UVIS2   & 800         & F555W   &14763     \\
ID5P01030     &2017-08-08              &WFC3 UVIS2   & 900         & F814W   &14763     \\
ID5P01040     &2017-08-08              &WFC3 UVIS2   &2920         & F631N   &14763     \\
ID5P01050     &2017-08-08              &WFC3 UVIS2   &2120         & F657N   &14763     \\
\enddata
\tablenotetext{a}{PI: A. Filippenko}
\tablenotetext{b}{PI: J. Maund}
\tablenotetext{c}{C. Conroy}
\end{deluxetable*}

%
%

\section{Results}
\label{sec:res}

%
%

\begin{deluxetable*}{llccccc}
    \tablecaption{Late-time HST photometry of SN 2011dh at UV and blue wavelengths and in narrow-band filters\label{tab:res:uvphot}}  
\tablehead{\colhead{MJD} & \colhead{Phase} & \colhead{F225W}&\colhead{F336W} & \colhead{F435W} & \colhead{F631N} & \colhead{F657N} } 
\startdata
56879.15       &  1166.65      	& 24.610 (0.111)& 24.801(0.105) & 24.616(0.038) & $\cdots$ & $\cdots$ \\
57388.05	   &  1675.55		& $\cdots$ 		& 24.471(0.125) &  $\cdots$ 	& $\cdots$ & $\cdots$\\
57973.14	   &  2260.64		& $\cdots$		& 24.827(0.084) &  $\cdots$		& 23.749(0.139) & 23.772(0.083)\\
\enddata
\end{deluxetable*}

\begin{figure}
\begin{center}
\includegraphics[width=0.5\textwidth]{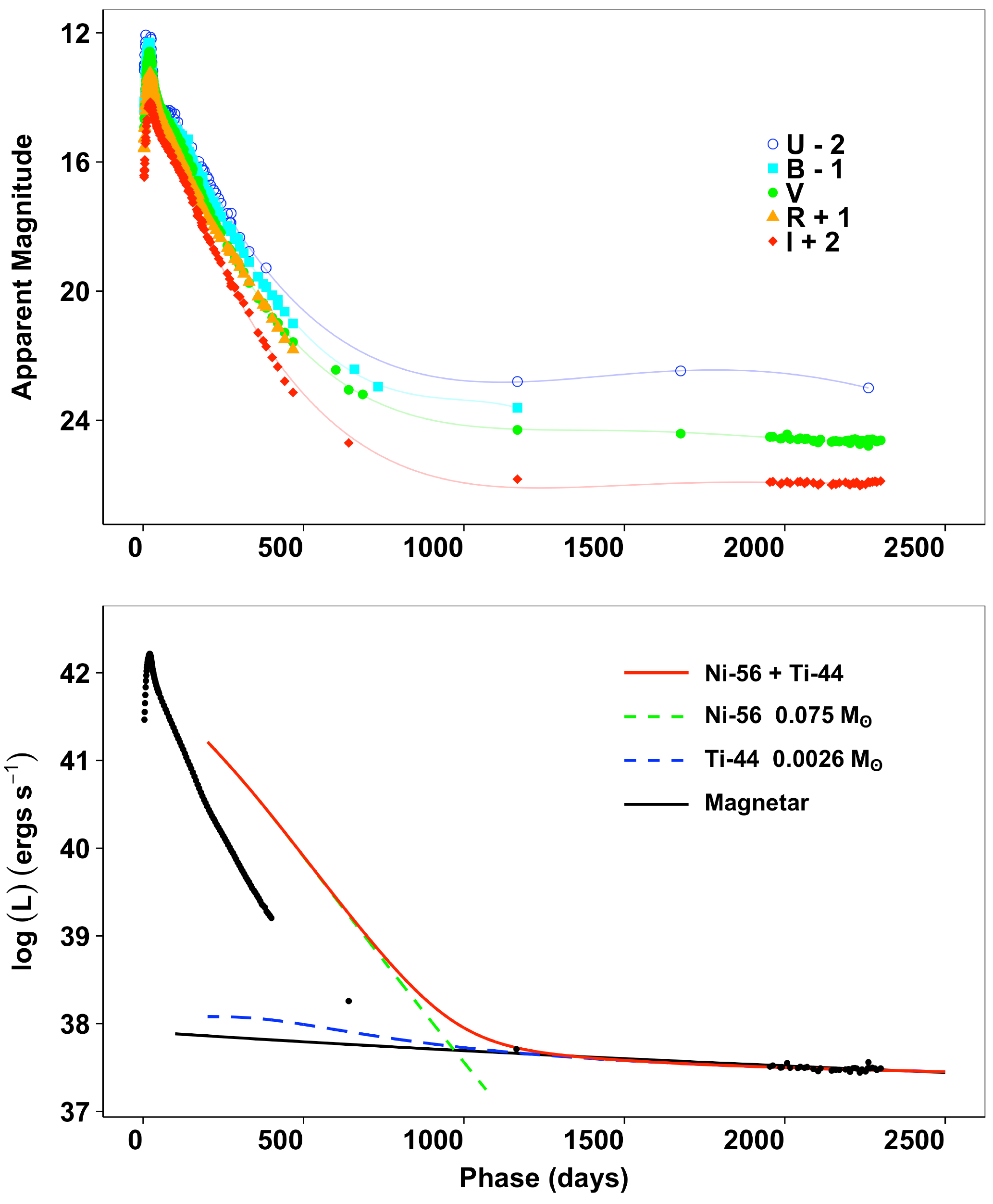}
\end{center}
\caption{The late-time light curve of SN~2011dh.  {\it (Top)}, The evolution of the optical light curve (for epochs $t > 640\mathrm{d}$ we adopt the photometry in the equivalent HST filters).  A polynomial fit to the data is shown by solid lines to indicate the approximate evolution in each filter band. {\it (Bottom)}, The evolution of the bolometric light curve.  For epochs $t < 640\mathrm{d}$ we adopt previously reported values of the luminosity \citep{2014A&A...562A..17E}, while for later observations the luminosities were derived using the HST equivalent of the $V-I$ colour, assuming a reddening of $E(B-V) = 0.07^{+0.07}_{-0.04}\,\mathrm{mags}$.  Overplotted are fits for the lightcurve decays expected for the decay chains due to $\mathrm{^{56}Ni}$ \citep[adopting the Ni mass derived by ][]{2015A&A...580A.142E} and $\mathrm{^{44}Ti}$ derived here \citep{2014ApJ...792...10S} and the luminosity evolution expected for a magnetar \citep{2010ApJ...717..245K}.}
\label{fig:latelc}
\end{figure}

\begin{deluxetable*}{llccc}
    \tablecaption{Late-time HST photometry of SN 2011dh at visible and near-infrared wavelengths\label{tab:res:visirphot}}  
\tablehead{\colhead{MJD} & \colhead{Phase} & \colhead{F555W}&\colhead{F606W} & \colhead{F814W}} 
\startdata
56353.43 & 640.93  & 23.057(0.018) & $\cdots$ & 22.703(0.026)\\
56879.15 & 1166.65 & 24.297(0.029) & $\cdots$ & 23.826(0.020) \\
57388.05 & 1675.55 & 24.415(0.037) & $\cdots$ & $\cdots$ \\
57666.23 & 1953.73 & $\cdots$ & 24.517(0.018) & 23.919(0.022) \\ 
57675.24 & 1962.74 & $\cdots$ & 24.513(0.017) & 23.901(0.020) \\ 
57699.07 & 1986.57 & $\cdots$ & 24.573(0.017) & 23.978(0.020) \\ 
57703.11 & 1990.61 & $\cdots$ & 24.561(0.017) & 23.952(0.020) \\ 
57719.68 & 2007.18 & $\cdots$ & 24.438(0.022) & 23.906(0.020) \\ 
57729.02 & 2016.52 & $\cdots$ & 24.579(0.018) & 23.953(0.021) \\ 
57753.06 & 2040.56 & $\cdots$ & 24.591(0.017) & 23.914(0.019) \\ 
57760.61 & 2048.11 & $\cdots$ & 24.561(0.017) & 23.897(0.019) \\ 
57773.93 & 2061.43 & $\cdots$ & 24.584(0.018) & 23.961(0.020) \\ 
57782.02 & 2069.52 & $\cdots$ & 24.576(0.017) & 23.904(0.019) \\ 
57801.74 & 2089.24 & $\cdots$ & 24.624(0.023) & 23.938(0.020) \\ 
57816.39 & 2103.89 & $\cdots$ & 24.688(0.020) & 24.003(0.023) \\ 
57823.61 & 2111.11 & $\cdots$ & 24.597(0.019) & 23.954(0.022) \\ 
57858.50 & 2146.00 & $\cdots$ & 24.670(0.016) & 24.004(0.016) \\ 
57863.53 & 2151.03 & $\cdots$ & 24.647(0.016) & 23.960(0.018) \\ 
57871.81 & 2159.31 & $\cdots$ & 24.643(0.012) & 23.985(0.014) \\ 
57882.36 & 2169.86 & $\cdots$ & 24.660(0.018) & 23.946(0.020) \\ 
57901.69 & 2189.19 & $\cdots$ & 24.644(0.013) & 23.925(0.015) \\ 
57911.22 & 2198.72 & $\cdots$ & 24.623(0.013) & 23.972(0.014) \\ 
57915.85 & 2203.35 & $\cdots$ & 24.710(0.014) & 24.000(0.015) \\ 
57925.59 & 2213.09 & $\cdots$ & 24.592(0.017) & 23.946(0.020) \\ 
57931.35 & 2218.85 & $\cdots$ & 24.584(0.018) & 23.959(0.021) \\ 
57936.62 & 2224.12 & $\cdots$ & 24.606(0.014) & 23.920(0.014) \\ 
57946.46 & 2233.96 & $\cdots$ & 24.741(0.017) & 24.027(0.020) \\ 
57953.15 & 2240.65 & $\cdots$ & 24.632(0.022) & 23.984(0.027) \\ 
57964.27 & 2251.77 & $\cdots$ & 24.706(0.019) & 23.990(0.023) \\ 
57971.62 & 2259.12 & $\cdots$ & 24.600(0.018) & 23.918(0.020) \\ 
57973.14 & 2260.64 & 24.688(0.039) & $\cdots$ & 23.812(0.045)\\
57978.10 & 2265.60 & $\cdots$ & 24.603(0.013) & 23.933(0.017) \\ 
57985.55 & 2273.05 & $\cdots$ & 24.588(0.013) & 23.907(0.015) \\ 
57993.49 & 2280.99 & $\cdots$ & 24.626(0.016) & 23.888(0.018) \\ 
57999.31 & 2286.81 & $\cdots$ & 24.657(0.020) & 23.916(0.024) \\ 
58011.36 & 2298.86 & $\cdots$ & 24.620(0.019) & 23.884(0.020) \\
\enddata
\end{deluxetable*}
The photometry of SN~2011dh from the late-time HST observation are presented for Ultraviolet (UV), blue and narrow-band filters in Table \ref{tab:res:uvphot} and for optical and near-infrared filters in Table \ref{tab:res:visirphot}.  The late-time light curve of SN~2011dh is presented, in combination with early-time observations (in comparable filters) from \citet{2014A&A...562A..17E} and \citet{2015A&A...580A.142E}, in Figure \ref{fig:latelc}.
After the steep decline at early times, $\sim \mathrm{500d}$ after explosion, the light curve of SN~2011dh exhibits a plateau at $\sim 1000\mathrm{d}$ and, at blue wavelengths, may exhibit a brief phase of rebrightening at $\mathrm{\sim 1675d}$ (however, due to the paucity of photometric coverage at these wavelengths, the significance of this rebrightening is debatable).  From $\mathrm{1950d}$, the light curve is seen to decline again and become redder in the optical bands with decline rates of $0.031 (\pm 0.008)$ and $0.0 (\pm 0.007)$ $\mathrm{mags\,{100d}^{-1}}$ in the HST $F555W/F606W$ and $F814W$ filters, respectively. 
The origin of the apparent $2\sigma$ deviation between the ACS and WFC3 $F814W$ photometry at 2259.12 and 2260.64d is unclear, especially given the similarities of the two $F814W$ filters; fluctuations in the photometry of this scale are observed elsewhere in the lightcurve, but not at successive epochs.  Investigations with DOLPHOT do not show any systematics differences between the photometry for other sources shared between the two sets of imaging.

\subsection{Bolometric Light Curve}
\label{sec:res:bolometric}
At early times ($t < 600\mathrm{d}$), the evolution of the bolometric luminosity has been previously presented by \citet{2015A&A...580A.142E}.  These measurements are based on integration over multi-wavelength observations, from the ultraviolet to the mid-infrared.  The majority of the late-time data, however, are composed of pairs of $F606W$ and $F814W$ ACS observations.  The determination of the bolometric luminosity is, therefore, limited by the colour information contained in these observations \citep[and is limited by the assumption that the overall shape of the SN SED is not affected by strong emission lines, which may be a reasonable assumption given the spectrum observed at 678d by][]{2013MNRAS.436.3614S}.  For the late-time data we apply fits to the Blackbody function, assuming a value for the reddening of $E(B-V) = 0.07^{+0.07}_{-0.04}\,\mathrm{mags}$ \citep{2014A&A...562A..17E}, to derive estimates for the bolometric luminosity.  Given the continuing colour evolution of the SN (see Fig. \ref{fig:latelc} and Table \ref{tab:res:visirphot}), the adoption of observations with a single filter and assuming a constant bolometric correction would be inappropriate.  For the multi-wavelength late-time datasets acquired in 2014 and 2017 (see Tables \ref{tab:res:uvphot} and \ref{tab:res:visirphot}), we find that the inclusion of additional photometry at other wavelengths does not significantly alter the derived luminosity ($\Delta \log L < 0.01\,\mathrm{dex}$) compared to using just $\sim V$ and $~I$ band photometry.  The error on the derived luminosity, at each epoch, is dominated by the uncertainty on the distance with $\Delta \left(\log (L/L_{\odot}) \right) \sim 0.11$.  A key parameter of interest is the rate of the decline of the bolometric light curve, for comparison with the expected temporal evolution of different mechanisms; therefore, we do not incorporate uncertainties on the distance modulus which would induce correlated errors between late-time data points.  We measure the decline of the bolometric luminosity for $t > 1950\mathrm{d}$ to be $-0.009 \pm 0.003\,\mathrm{dex}\,100\mathrm{d}^{-1}$.

\section{Analysis}
\label{sec:ana}
We consider a number of possible origins for the brightness of the source at the position of SN~2011dh: a remaining stellar object (a possible companion), circumstellar interaction, the decay of radioactive elements and a light echo. 

\subsection{Stellar residual}
The evolution of the late-time light curve across all wavelengths, despite changing slowly, is significant and is inconsistent with the constant brightness expected if the source at the SN location is a non-varying star.  As noted by \citet{2015MNRAS.454.2580M} the spectral energy distribution (SED) of the late-time source observed in 2014 at the position of SN~2011dh is inconsistent with stellar SEDs.

\subsection{Circumstellar interaction}
\label{sec:res:csm}
\begin{figure}
\begin{center}
\includegraphics[width=0.8\linewidth]{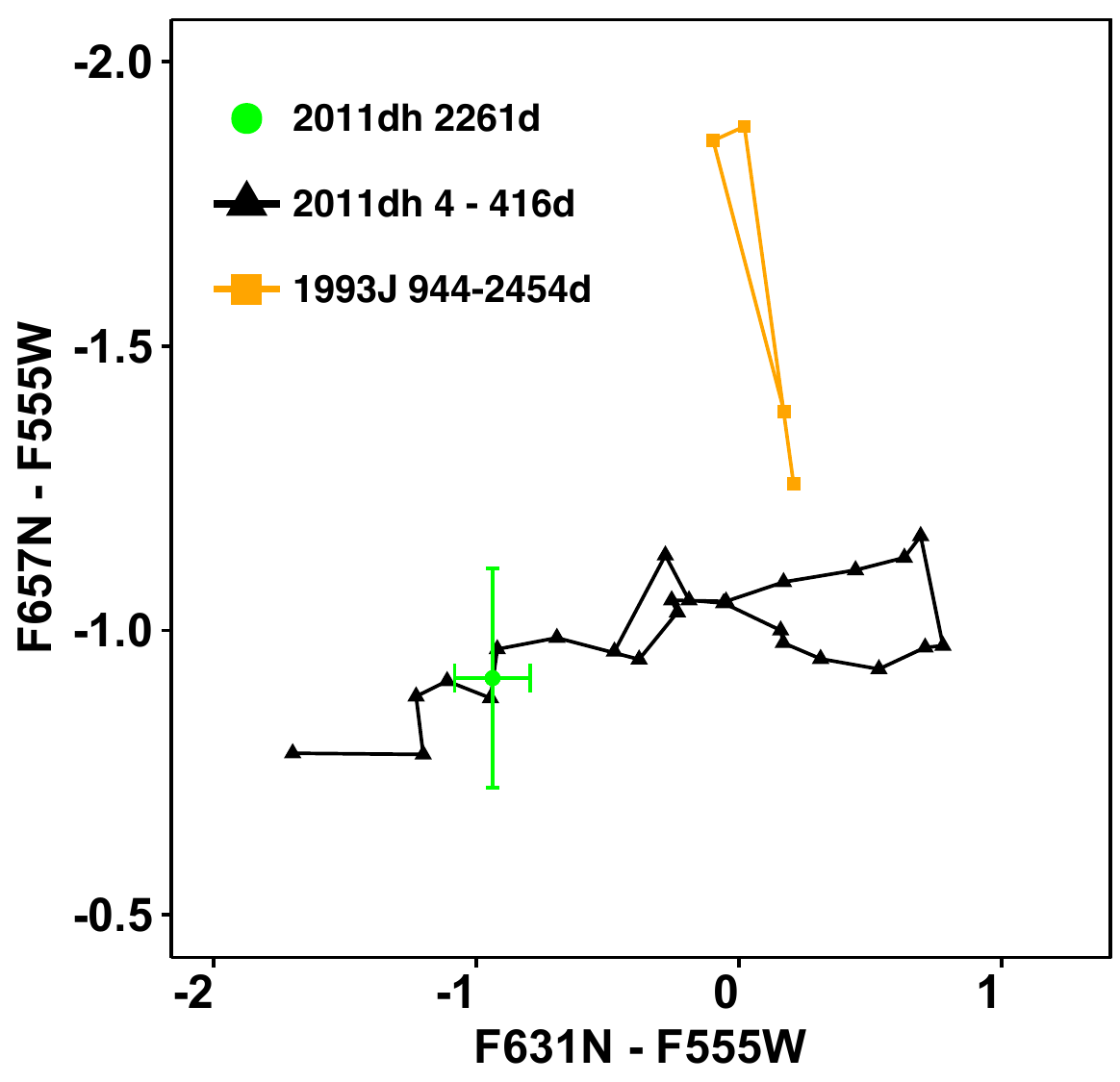}
\end{center}
\caption{Narrow-band direct and synthetic photometry of SNe 2011dh and 1993J for $\mathrm{[O I]}$ $\lambda 6300$ and $\mathrm{H\alpha}$ $\lambda 6563$, relative to broad band $\sim V$ ($F555W$) photometry (as an estimate of the “continuum”).  In black are shown synthetic photometry of early spectra of SN 2011dh and the orange points indicate the synthetic photometry of late-time spectra of SN 1993J \citep{2014AJ....147...99M}.  SN 1993J, which has been established to be undergoing significant CSM interaction, shows a clear excess of $\mathrm{H\alpha}$ flux (corresponding to broad box-like profiles observed in the spectrum) which is not seen in the colours of SN 2011dh at similar epochs.}
\label{fig:res:csm}
\end{figure}

An alternative source of luminosity could be the reinvigoration of interaction of the ejecta with surrounding circumstellar material (CSM).  In spectroscopic observations at 628 and 678 days, there was no evidence for strong emission lines (in particular, box-like spectral line profiles) consistent with CSM interaction at that time \citep{2013MNRAS.436.3614S}.  We compared our photometry from the 2017 multi-colour HST observations with synthetic photometry of late-time spectra of SN 1993J, confirmed to be undergoing significant CSM interaction, at $\mathrm{2453d}$ post-explosion \citep{2014AJ....147...99M}\footnote{Retrieved from the WISeREP archive - https://wiserep.weizmann.ac.il - \citet{2012PASP..124..668Y}}.  The magnitude $m$ that would be observed given a particular bandpass $P_{\lambda}$ for a target spectrum with flux density $f_{\lambda}$, is calculated as;
\begin{equation}
m = -2.5 \log_{10} \left( \frac{\int^{\lambda_{2}}_{\lambda_{1}}\lambda f_{\lambda} P_{\lambda} \mathrm{d}\lambda}{\int^{\lambda_{2}}_{\lambda_{1}}\lambda f^{0}_{\lambda} P_{\lambda} \mathrm{d}\lambda} \right) +  m^{0}_{P}
\label{eqn:synphot}
\end{equation}
where $f^{0}_{\lambda}$ is the flux density, as a function of wavelength, of a reference object \citep[e.g. Vega;][]{2004AJ....127.3508B} and $m^{0}_{P}$ is an additional zeropoint offset for defining the photometric system (for HST filters this is defined as exactly $m^{0}_{P} = 0$).  To calculate photometry using the HST filter set, we utilised tabulated filter transmission functions (and throughputs for the HST optical train)\footnote{http://www.stsci.edu/hst/observatory/crds/throughput.html}.  We specifically compared our narrow band ($F631N$ and $F657N$) photometry in which strong emission lines of $[\mathrm{O I}]\,\lambda 6300$ and $\mathrm{H\alpha}$, if present, would cause significant colour excesses compared to corresponding photometry of the SN ``continuum" (here defined as the the broad-band $F555W$ photometry).  As shown in Figure \ref{fig:res:csm},  SN~2011dh does not exhibit a significant excess of $\mathrm{H\alpha}$ emission in 2017, but rather shows colours similar to earlier spectra at $\sim 400\mathrm{d}$ \citep{2015A&A...580A.142E} and $628 - 678\mathrm{d}$, such that there is no evidence that the optical spectrum of SN~2011dh is powered by CSM interaction that commenced since the last spectroscopic observations.

\subsection{Magnetar}
If the ejecta are optically thin, such that the diffusion time scale is short, a highly magnetized ($\sim (3.4 \pm 0.6)\times 10^{15}G$), slowly rotating (initial period $0.92\pm0.11\,\mathrm{s}$) magnetar \citep{2010ApJ...717..245K,2010ApJ...719L.204W} could replicate the late-time ($\mathrm{>1950d}$) luminosity; however, the decline observed at $>1950\mathrm{d}$ (see Figure \ref{fig:latelc}) is shallower than predicted by the magnetar model.

\subsection{Radioactive Nuclides}
The decline of the $\sim V$ and $I$ band light curves, reported in Section \ref{sec:res} is much shallower than the expected decline if the SN were powered by just the decay or $\mathrm{^{56}Ni}$.  The decay of the daughter product $\mathrm{^{56}Co}$ to $\mathrm{^{56}Fe}$, relevant at later epochs, would give a corresponding decline rate of $0.97\,\mathrm{mags}\,100\mathrm{d}^{-1}$, given a half-life of $t_{\mathrm{1/2}}=77.2\mathrm{d}$.  

The mass of radioactive $\mathrm{^{56}Ni}$ produced in the explosion was well constrained at early-times to be $M_{Ni} \sim 0.075M_{\odot}$ \citep{2012ApJ...757...31B,2015A&A...580A.142E,2013MNRAS.436.3614S}, with increasing transparency to $\gamma$-rays causing a faster decline than expected.  For the late-time photometry, when the ejecta is more likely to be optically thin, we use the technique of \citet{2014ApJ...792...10S} to estimate the quantities of different radioactive elements required to reproduce the late-time bolometric luminosity.  In addition to $\mathrm{^{56}Ni}$, we also consider the roles of $\mathrm{^{57}Ni}$ and $\mathrm{^{44}Ti}$.  In order to approximate the flattened late-time bolometric light curve at $t > 1000\mathrm{d}$ using radioactive species would necessitate the presence $(2.7 \pm 0.5) \times 10^{-3}M_{\odot}$ of $\mathrm{^{44}Ti}$, but negligible quantities of $\mathrm{^{57}Ni}$ ($<3 \times 10^{-5}M_{\odot}$).  At such late-times, changing the degree of transparency to $\gamma$-rays has little effect on the rate of the decline of the light curve (although, in our calculation, we assume full lepton trapping).

\subsection{Light Echo}
Light echoes can arise from light originating from SNe at early times being scattered into the line-of-sight by nearby dust \citep{1939AnAp....2..271C}.  As the scattered light has taken a longer path to reach the observer, there is a delay between the arrival of these photons and those photons that have travelled directly from the SN.  Such echoes are generally discernible, years post-explosion, as spatially resolved arcs or rings around SNe \citep[e.g. the Type IIb SN 1993J, for which dust in a foreground sheet produced spatially resolved arcs around the SN;][]{2003ApJ...582..919L, maund93j}.  In the case of SN~2011dh, there is no evidence for spatially extended arc-like structures around the SN in the late-time 2017 observations (see Fig. \ref{fig:picture}).

\begin{figure}
\begin{centering}
\end{centering}
\includegraphics[width=8cm]{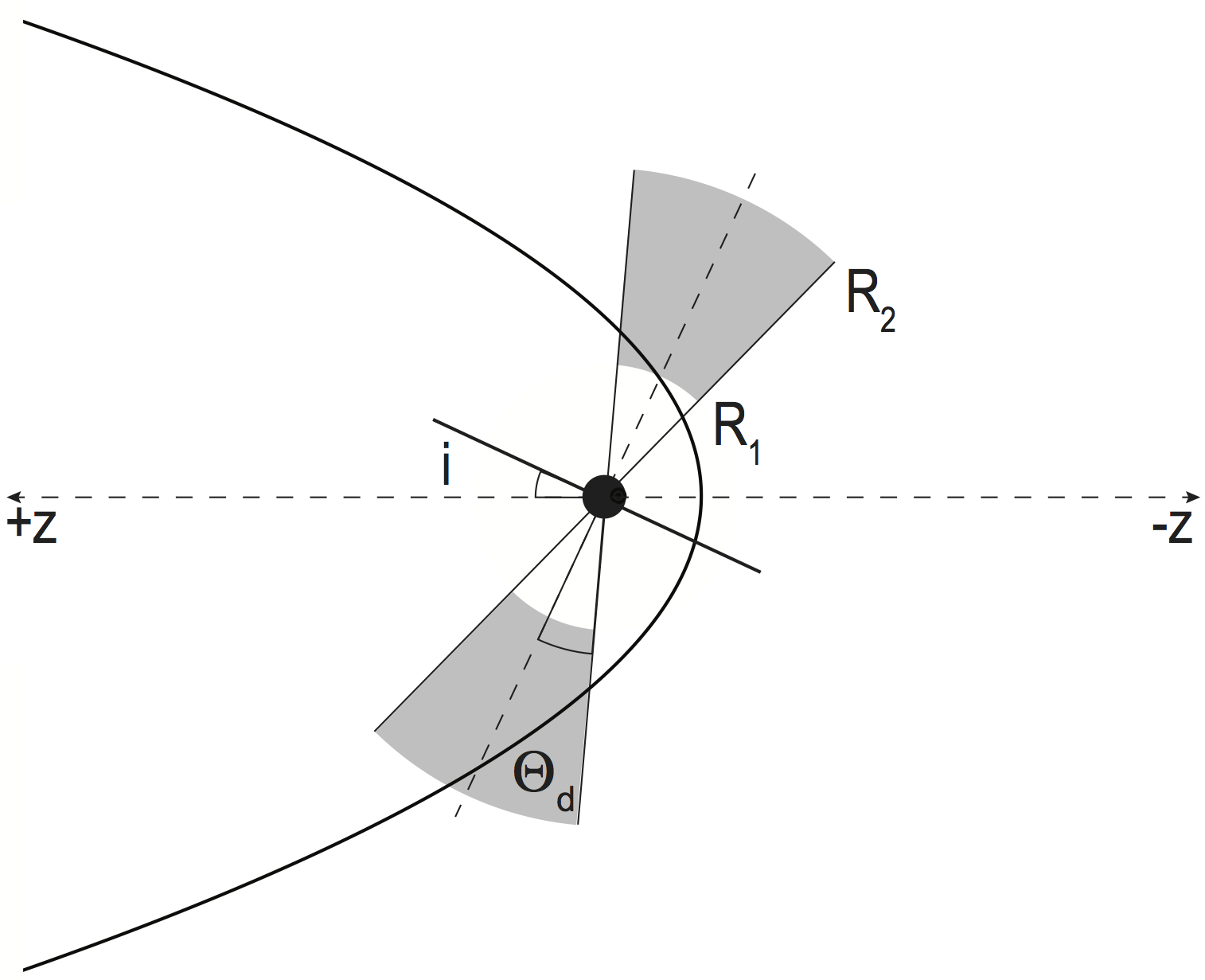}
\caption{A cross-sectional schematic of the light echo formation scenario in a flared disk geometry of dust, inclined at an angle $i$ with respect to the line-of-sight. The grey shaded area indicates the location of the dust (with respect to the SN, indicated by the black point). The observer is located to the left of the diagram. The parabola indicates an ``iso-delay surface". For light originating from the SN that scatters on dust lying on this parabola, the light will arrive at the observer with the same apparent time delay.}
\label{fig:ana:disk}
\end{figure}

The brightness of a light echo depends on the scattering properties of the dust and the geometry of the distribution of the dust around the SN, which also dictates the time delay experienced by the light and the evolution of the observable light curve.  A Monte-Carlo scattering model \citep{1977ApJS...35....1W,2005MNRAS.357.1161P} was developed to explore the complex geometric distributions that might be responsible for an unresolved or compact light echo, consistent with the source recovered at the SN location in the 2017 observations.   We utilised the dusting scattering properties of albedo $\omega$, extinction and scattering cross-sections $C_{ext}$ and $C_{sca}$ per H atom, and the scattering asymmetry parameter $g = <\cos \theta>$ appropriate for Galactic-type dust presented by \citet{2001ApJ...548..296W}\footnote{https://www.astro.princeton.edu/$\sim$draine/dust/scat.html}.  To improve the efficiency of the calculation, each photon was forced to scatter once and at each scattering event the new direction of travel of the photon was determined using the Henyey-Greenstein phase function \citep{1941ApJ....93...70H}.  Values of the dust parameters at the wavelengths of the relevant filters were derived by interpolating the tabulated values of the dust properties.  We considered three families of geometric configurations for the dust: 1) plane sheet of dust with thickness $\Delta R$ offset from the SN by a distance $R$ with constant density $n_{0}(H)$; 2) a spherical shell centred on the SN, with inner and outer radii $R_{1}$ and $R_{2}$ and density for $R_{1} \le r \le R_{2} = n_{0}(H)\left(r / R_{1}\right)^{-\alpha}$, which for a stellar wind $\alpha = 2$ and $n_{0}(H)$ is the reference density of hydrogen ($\mathrm{cm^{-3}}$); and 3) a flared disk with inner and outer radii $R_{1}$ and $R_{2}$, inclined by an angle $i$ to the line-of-sight, and flare angle $\Theta_{d}$ such that at a distance $d$, in the plane of the disk, the thickness of the disk is $\Delta z = 2  d\tan(\Theta_{d})$ (see Figure. \ref{fig:ana:disk}).

In order to model the light incident on the dust grains we considered the reflection of light arising from the early ($<50\mathrm{d}$) lightcurve of SN~2011dh.  The early-time lightcurve was reconstructed from the early observations using combinations of the broad-band photometry and spectroscopy \citep{2014A&A...562A..17E} to determine the flux in the relevant HST filters of the late-time observations.  The original optical data, from ground-based observations, were corrected to the standard Johnson-Cousins system.  We used an interpolation technique \citep[Gaussian processes;][]{rgpfit} to calculate the brightness of the SN at intermediate epochs (with daily sampling), covering the wavelength range spanning from the Swift telescope observations in the UV, the ground-based Johnson-Cousins optical observations and infrared observations extending to the $K$-band.  For each epoch an SED was calculated, and broad-band photometry in the corresponding HST filter set was calculated using linear interpolation of the SED.  In general, given the close proximity of the HST filter set to the standard filters, we found differences between the observed and interpolated photometry of $<0.05\,\mathrm{mags}$ in the standard optical filter sets.  This calculation was then cross-checked against synthetic photometry (see Equation \ref{eqn:synphot}), using filter band-passes appropriate for the HST filter system, of the sparser collection of published flux-calibrated early-time spectra of SN~2011dh \citep{2014A&A...562A..17E}.  These indicated that the interpolated early-time $F606W$ light curve required a constant adjustment of $+0.05\,\mathrm{mags}$, to incorporate the effects of emission lines not included in the SED constructed from ground-based broad-band photometry.

The early-time spectra were crucial for the determination of the lightcurves appropriate for the narrow-band filters ($F631N$ and $F657N$) used in the 2017 observations.  As these filters are more susceptible to line features that might occur within the corresponding bandpass, we found synthetic photometry of the early-time spectra to yield fluxes significantly discrepant from those derived using the SED interpolation scheme.  The interpolation of the SED was found to significantly over-estimate the brightness in the $F631N$ bandpass, whilst underestimating the brightness in the $F657N$ bandpass (due to the strong absorption and relatively weak emission due to $\mathrm{H\alpha}$ occurring in these two wavelength ranges in the early spectra).  Our reconstruction of the early time light curves in the two ultraviolet ($F225W$ and $F336W$) filters, with WFC3 UVIS, are important exceptions, as the wavelength ranges are partially or completely unobserved in ground-based photometry and spectroscopy due to the atmospheric cutoff in the UV.  We have considered the reported Swift UV and $UBV$ photometry \citep{2014A&A...562A..17E} and used an interpolation scheme to calculate the corresponding fluxes in the HST filters.

As a final test of the reconstruction of the light curves in all the non-standard HST filters, we utilised an HST Space Telescope Imaging Spectrograph (STIS) spectrum of SN 2011dh from 2011 Jun 24.13 UT (at 23.6 days after explosion, for program GO-12540, PI Kirshner), extending from $1663 - 10240 \mathrm{\AA}$, to establish that the systematic errors associated with the interpolated and synthetic photometry are, at most, $0.1\,\mathrm{mags}$ (and, generally, better than this for HST filters that closely resemble their ground-based Johnson-Cousins equivalents).

Suitable light echo models are required to fit the overall time evolution of the brightness and colour of the light curve observed at late-times; in particular,  the onset of the flattening of the light curve at $t > 500\mathrm{d}$, the possible bump in the light curve at $\sim 1700\mathrm{d}$, the continued presence of the light echo at $2300\mathrm{d}$ and the limits on the amount of foreground extinction from the reddening derived toward the SN itself \citep{2014A&A...562A..17E}.  

The identification of an appropriate geometric configuration for the scattering dust is non-trivial: 1) the response of the light echo brightness to changes in the parameters, such as the dust density, is not linear; 2) the assumed geometries may be simplifications of more complicated structures around SN~2011dh; and 3) the available photometry does not, necessarily, have the diagnostic capability to identify a unique model.  We have, therefore, made a survey of likely families of models for comparison with the observations.   In the first instance, we only considered fits to the comprehensive set of late-time $F555W/F606W$ and $F814W$ observations, and then used the observations at other wavelengths to test the overall predictive quality of the models.

\begin{figure}
\begin{center}
\includegraphics[width = 8cm]{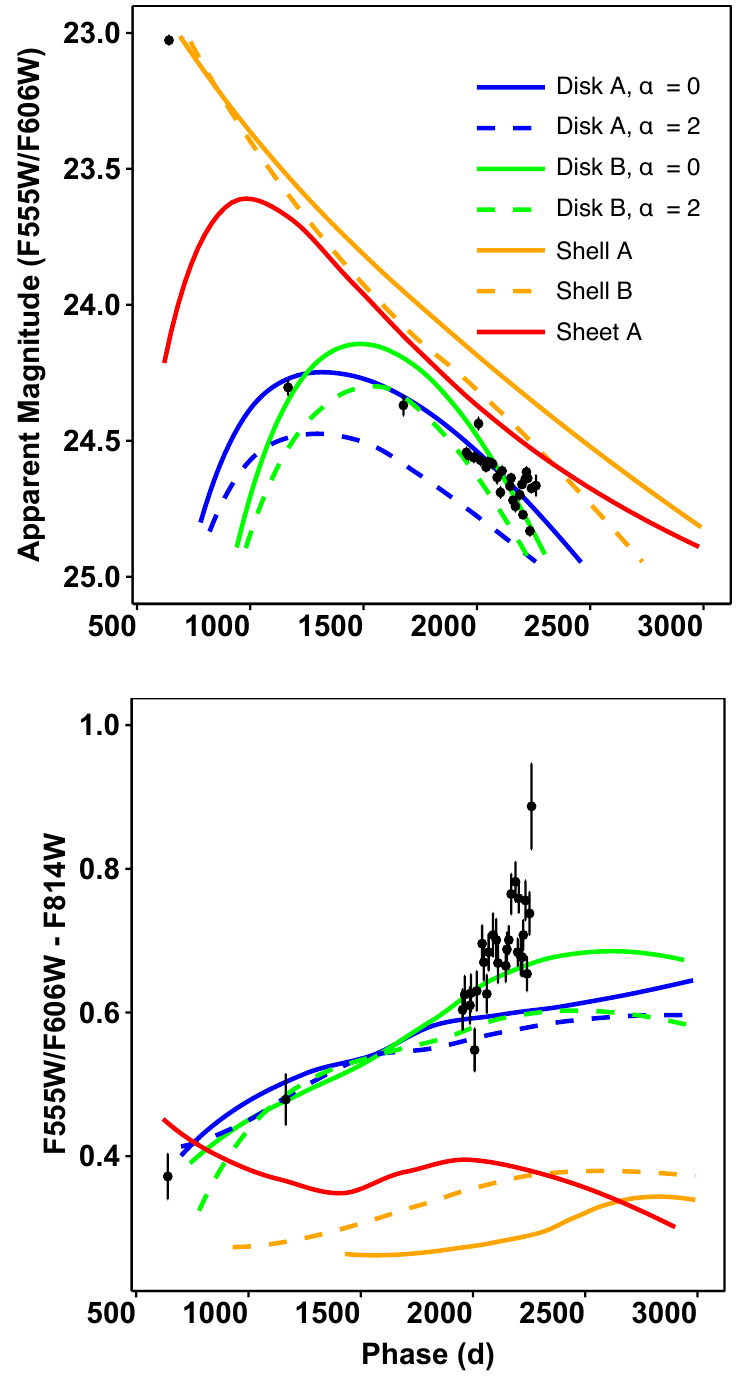}
\end{center}
\caption{Predicted late-time $F555W/F606W$-band light curve and $F555W/F606W - F814W$ colour for seven different dust geometries:  disk A ($n_{0}(H) = 250\,\mathrm{cm^{-3}}$,$R_{1} = 1.8$,$R_{2} = 2.6\,\mathrm{pc}$, $\Theta_{d} = 17.5^{\circ}$) for $\alpha = 0$ and $2$; disk B ($250$,$2.1$, $2.7$, $17.5$) for $\alpha = 0$ and $2$; Shell A ($250$,$2.0$, $2.5$) for $\alpha = 0$; Shell B ($250$,$1.0$, $2.0$) for $\alpha = 2$; and Sheet A ($n_{0}(H) = 500\,\mathrm{cm^{-3}}$,$R = -0.2\,\mathrm{pc}$, $\Delta R = 0.2\,\mathrm{pc}$, $\alpha = 0$).}
\label{fig:pred}
\end{figure}

We find that continuous dust sheets and spherical shells do not yield light curves that exhibit the appropriate degree of flattening/rebrightening and do not predict the appropriate colour evolution (see Fig. \ref{fig:pred}).  Foreground and background dust sheets (at all inclinations) that can predict the decline rate at $\sim 2000\mathrm{d}$, overpredict the brightness at $\sim 1000\mathrm{d}$.  For hydrogen densities of $n_{0}(H) = 100\,\mathrm{cm^{-3}}$, a background sheet with $R = -0.2$ and $\Delta R = 0.4\,\mathrm{pc}$ yields $\chi^{2} = 3538$ (44 d.o.f.), if the sheet is observed with an inclination of $50^{\circ}$. For both dust sheets and spherical shells, scattering of early-time light by dust in front of the SN produces a significant light echo contribution at early times, that has been ruled out by analyses of the early-time light curve ($<\mathrm{600d}$) \citep{2015A&A...580A.142E}.  This effect is amplified by the preference for forward scattering (as parameterised by the scattering asymmetry parameter $g$) such that, even in the case of a spherical shell, the light echo produced by the near side of the shell is much brighter than that produced at later times when photons must be scattered through $\sim 180^{\circ}$ on the far side of the shell.

As shown in Fig. \ref{fig:pred}, a disk-like geometry, approximately in the plane of the sky, can qualitatively reproduce the brightness profile and the increasing red colour observed for SN~2011dh at late-times. The colour evolution at $\sim 2000\mathrm{d}$ is particularly crucial to assessing the appropriateness of the geometry of the dust distribution: the density needs to be sufficiently high to reproduce the brightness of the dust echo and segregate the blue and red photons to a sufficient degree to yield the progressively redder colour observed at later epochs.  For a disk-like geometry, oriented close to the plane of the sky, the optical depth from the SN to the point of scattering (in the plane of the disk) may be high (inducing the observed colour) but the optical depth to reach the observer after scattering, perpendicular to the plane of the sky, is low allowing photons to escape without being subject to additional extinction.  The latest data ($>1950\mathrm{d}$) are insufficient for constraining the inclination of the disk. Incorporating the earlier observations limits the inclination $<20^{\circ}$ with disks observed edge-on over predicting the brightness at $\sim 1000\mathrm{d}$, for the same reasons as the foreground sheets and the spherical shells.  The orientation of the disk configuration almost in the plane of the sky is important for reproducing the observed flattening/rebrightening and gentle fall off of the late-time light curve.  The characteristic density for the disk is $\sim 250 - 275\,\mathrm{cm^{-3}}$, and to reproduce the time evolution of the brightness requires the disk to lie within $\sim 1.8$ to $\sim 2.7\,\mathrm{pc}$ of the SN, with the flare angle of the disk in the range $\Theta_{d} = 15 - 22.5^{\circ}$.  There is no preference for a disk of either constant density or one with a fall-off consistent with a stellar wind ($\propto r^{-2}$).    The large parameter space for possible geometries that may be responsible for such a light echo (which may not be well described by discrete boundaries), as well as the large range of possible dust grain properties, mean that identifying a precise fit to the observed data is an extremely computationally expensive problem, and beyond the scope of this work.

\begin{figure*}
\begin{center}
\includegraphics[width = 9cm,angle=90]{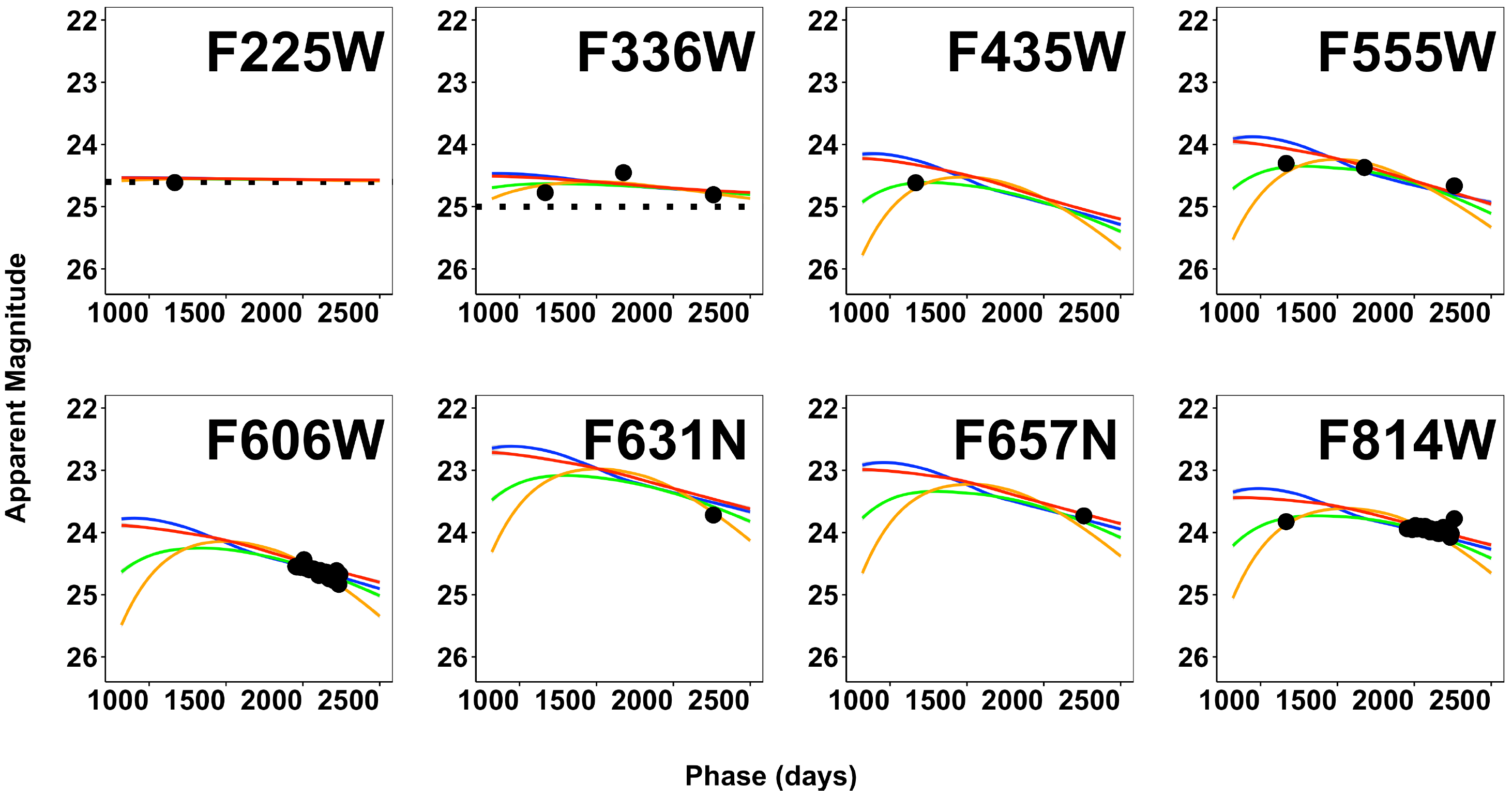}
\end{center}
\caption{Multiwavelength late-time light curves for SN~2011dh.  Overlaid are four models for light echoes originating in a disk-like medium around SN~2011dh, with parameters $\{ n_{0}(H) = 250\,\mathrm{cm^{-3}}$, $R_{1} = 1.8$, $R_{2} = 2.6\,\mathrm{pc}$ , $\alpha = 0$, $\Theta_{d} = 17.5^{\circ}, \cos i = 0.875 \}$ (blue; $\chi^{2} = 58.1$ for 47 d.o.f.), $\{250, 2.1, 2.7, 0, 17.5, 0.925 \}$ (green; $\chi^{2} = 42.3$ for 47 d.o.f.), $\{275, 2.1, 2.5, 0, 22.5, 0.975 \}$ (orange; $\chi^{2} = 59.5$ for 47 d.o.f.) and $\{275, 2.1, 2.7, 0, 22.5, 0.875 \}$ (red; $\chi^{2} = 53.2$ for 47 d.o.f.) with a constant flux component (black dotted line), corresponding to the binary companion, added in the $F225W$ and $F336W$ bands.}
\label{fig:panel}
\end{figure*}

\begin{figure}
\begin{center}
\includegraphics[width = 8cm]{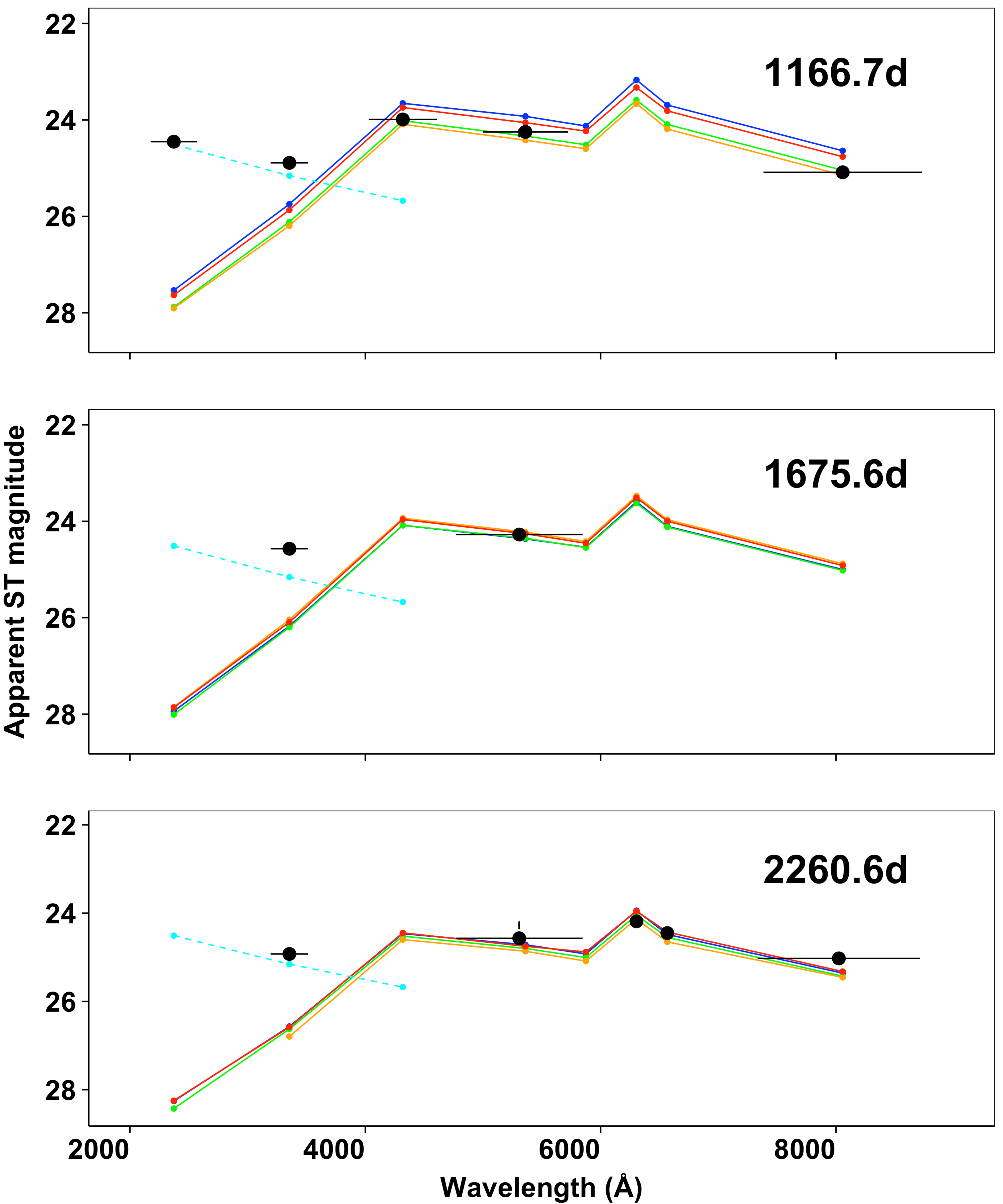}
\end{center}
\caption{Observed and predicted SEDs for the late-time HST observations of SN~2011dh.  The SEDs are presented in ST magnitudes (for which, in Equation \ref{eqn:synphot}, $f^{0}_{\lambda} = 3.63 \times 10^{-9}\,\mathrm{ergs\,cm^{-2}\,s^{-1}\,\AA^{-1}}$), which provide a closer representation of the underlying echo spectrum.  The light echo model SEDs are for the same models as those presented in Fig. \ref{fig:panel}.  The SED of the binary companion is presented by the dashed, cyan line.}
\label{fig:sed}
\end{figure}

Over all optical wavelengths these disk models yield good fits (see Figures \ref{fig:panel} and \ref{fig:sed}), however significant discrepancies are found at bluer wavelengths (corresponding to the $F225W$, $F336W$ and $F435W$ filters) for which the light echo model systematically under-predicts the observed brightness.  Utilising the light echo models derived independently for redder bands, the deficit in the blue bands can be modelled as a constant additive flux contribution across all epochs.  The flux deficit in the $F225W$ and $F336W$ bands (and limits in the $F435W$ band) constrain the brightness of this constant source to be $m_{F225W} = 24.67 \pm 0.06$, $m_{F336W} = 25.04 \pm 0.04$ and $m_{F435W} > 26.3\,\mathrm{mags}$ (see Figure \ref{fig:sed}).  

At the assumed distance of M51, an ACS/WFC pixel corresponds to $1.90\,\mathrm{pc}$.  The maximum spatial extent for a light echo arising from a disk of dust is $\sim 2 \times 2.7\,\mathrm{pc}$ for $i = 0^{\circ}$, corresponding to an annulus with diameter of $\sim 2.8$ pixels.  The observed Full Width at Half Maximum (FWHM) measured from the ACS observations (in the $F555W$ and $F606W$ bands at $0.05\,\arcsec\,\mathrm{px^{-1}}$), is $1.9\,\mathrm{px}$. An echo would, therefore, appear as a broadened profile compared to the Point Spread Function (PSF), with a flattened peak.  If the dust disk were inclined with respect to the line of sight, it would not produce a symmetric ring-like light echo.  Instead the projected geometry of the dust disk would be an ellipse, with major axis corresponding to the real diameter of the dust disk but with an axial ratio of  $\cos (i)$.   As shown in Fig. \ref{fig:ana:disk}, a further consequence of an inclined configuration is that the light echo would appear to first arise from dust on the near of the SN and then later on the opposite, far side of the SN.  For an inclined dust disk, the light echo would not appear to span the full spatial extent of the diameter of the disk.

In order to analyse the spatial characteristics of the late-time source, we consider three epochs of the ACS/WFC $F555W$ and $F606W$ at 1166.65, 1953.73 and 2298.86d.  These observations were specifically chosen as they were composed of multiple exposures, at multiple dither pointings, which could be used to facilitate the reconstruction of the images with better sampling with pixels of $0.025\,\arcsec$.  At the three epochs, the FWHM of the late-time source was measured to be $5.1$, $4.1$ and $5.2\,\mathrm{px}$, respectively.  For this imaging configuration, the empirical FWHM of the PSF was measured from isolated bright stars to be $3.0\,\mathrm{px}$.  The FWHM of the late-time source is $\sim 1 - 2 \, \mathrm{px}$ larger than the stellar PSF, implying that the source is clearly extended. 

From the light echo models, approximate measures of the expected extent for an observed echo were derived.  The coordinate of last scattering for each photon was used to reconstruct images of the evolving light echo to determine the relative brightness as projected on the plane-of-the-sky.  The predicted light echo was convolved with a bivariate normal distribution, to approximate the PSF, and then the FWHM in two orthogonal directions was calculated to infer the FWHM of the resulting source.  The models themselves predict, at $0.025 \arcsec\,\mathrm{px^{-1}}$, the FWHM of the late-time source should be $\sim 4 - 6\,\mathrm{px}$ across the three epochs.  More detailed inferences from these models are complicated by the relative faintness of the observed late-time source and the proximity of other nearby sources seen in the observations.  Although the models have been convolved with an approximation of the PSF, the distribution of flux as observed by ACS is itself pixellated.  Beyond demonstrating that the late-time source is extended, it is difficult to constrain the parameters of the disk by comparing the FWHM we measure from the convolved model predictions with the observed data, as the measured FWHM is also dependent on the position angle of the projected major axis of the dust disk on the plane of the sky relative to orientation of the pixel edges.  A more detailed analysis would require modelling of the subsampled HST/ACS PSF (including the complex role of changes in the HST focal length) and intricate scene modelling of the late-time source, the background and all the surrounding sources; which is beyond the scope of this paper.

\section{Discussion}

\subsection{Which energy source is responsible?}
Despite the usual considerations of the brightness evolution of SNe and their eventual disappearance over timescales following the decay of radioactive elements, there is an emerging picture, in particular for Type Ia SNe, that observations at later epochs may present a more complicated evolution \citep{2018ApJ...852...89Y,2018arXiv181007258G}

Based on the smooth evolution of the light curve, and the lack of a significant colour excess at the wavelengths associated with strong emission lines, we find no evidence to support the onset of interaction of the ejecta of SN~2011dh with a dense CSM after the last epochs of reported spectroscopy \citep{2013MNRAS.436.3614S}.  

From Figure \ref{fig:latelc}, it can be seen that both the decay of radioactive elements and the magnetar can provide reasonably good matches to the bolometric lightcurve.  As noted in Section \ref{sec:res:bolometric}, some caution is required in this analysis as, at best, the reconstructed luminosity is calculated using only broad-band photometry spanning from UV to near-infrared wavelengths; for the vast majority of the late-time data, the luminosity is constrained by photometry at only two wavelengths.  We find that the magnetar model can be fit to reproduce the luminosity gradient at $\sim 2000\mathrm{d}$, but this leads to an overprediction of the luminosity at $\sim 1000\mathrm{d}$.  Conversely, decay of radioactive nuclides can provide an excellent fit to the luminosity at $\sim 1000\mathrm{d}$ and reproduce the later decline rate.  This fact is presented in the difference between the Bayesian evidence \citep[the marginal likelihood;][]{2004AIPC..735..395S} of the fits to the two models, with $\Delta \ln B = +6.09$ that, following the scale of \citet{Jeffreys61}, constitutes very strong evidence in favour of radioactive decay being the energy source responsible.  We also note that, for the magnetar models, the implied initial spin period $P = 0.92\mathrm{s}$ is in the extreme tail of the neutron star initial spin period distribution presented by \citet{2013MNRAS.432..967I}.

For both the magnetar and radioactive decay scenarios, we have have to make assumptions about the fraction of energy that is captured by the ejecta and emitted as light.  For the magnetar model, we have assumed complete ``trapping", whereas by adopting the \citet{2014ApJ...792...10S} model for radioactive decay we have assumed a standard evolution of the $\gamma$-ray opacity, while assuming full lepton trapping.  In the latter case, there is evidence from early-time observations that this prescription may not be valid, with the light curve decaying faster than expected for $\mathrm{^{56}Ni}$ and $\mathrm{^{56}Co}$ (see Figure \ref{fig:latelc}) as evidence of the ejecta being optically thin to $\gamma$-rays and incomplete lepton trapping;  both \citet{2015A&A...580A.142E} and \citet{2013MNRAS.436.3614S} suggest increasing opacity and trapping can explain the light curve later becoming shallower at $\sim 300\mathrm{d}$.

The light echo model, perhaps, provides the most convincing case for the origin of the late-time brightness. Unlike the other energy sources, for which the luminosity rather than the SED is the key prediction, the light echo model must specifically address all the observables.  There are three major supporting cases for this model of the origin of the  late-time light curve evolution of SN~2011dh: 1) a light echo can predict the overall brightness evolution of the late-time light curve after $\mathrm{1000d}$ (see Fig. \ref{fig:panel}); 2) this light echo model can also reproduce the evolution of the wavelength dependence of the late-time light curve with time, which is primarily dependent on the early-time colour evolution of the SN at $t < \mathrm{50d}$ (see Fig. \ref{fig:sed}); and 3) the model predicts, except partially at the blue bump at $\mathrm{1675d}$, a brightness deficit consistent with a contribution from a constant brightness source (i.e. a binary companion).

\subsection{The importance of radioactive nuclides}
The amount of $\mathrm{^{44}Ti}$ inferred from the late-time bolometric light curve would require SN~2011dh to have produced a factor of $\sim 48$ times more titanium than produced in SN 1987A and $16$ times more than produced in the SN responsible for the Cas~A SN remnant \citep{2012Natur.490..373G,2014Natur.506..339G,2014ApJ...792...10S,2017ApJ...834...19G}.  This amount would make SN~2011dh an exceptional producer of $\mathrm{^{44}Ti}$, but it is also difficult to reconcile with the amount of $\mathrm{^{44}Ti}$ inferred from late-time spectra by \citet[][; $2.6\times 10^{-5}M_{\odot}$ for the $13M_{\odot}$ progenitor model]{2015A&A...573A..12J}, which is smaller by a factor of $\sim 100$.  Conversely, \citet{2006A&A...450.1037T} note that there is shortage of Galactic $\mathrm{^{44}Ti}$ $\gamma$-ray sources to explain the Galactic abundance of $\mathrm{^{44}Ca}$ and hypothesise that SNe responsible for the production of significant quantities of $\mathrm{^{44}Ti}$ may be atypical events (and, by extension, SN 2011dh could be one of these rare events).  
It should be noted that, although we are requiring late-time optical light curves to infer the presence of $\mathrm{^{44}Ti}$, the presence of this element can be directly observed through X-ray and $\gamma$-ray observations, in particular with the {\it NuSTAR} and {\it INTEGRAL} observatories.  Given the amount of $\mathrm{^{44}Ti}$ that might be suggested by the late-time optical light curve, following \citet{2012Natur.490..373G} and \citet{2015ApJ...814..132L} we estimate the corresponding flux arising from SN~2011dh in the $\mathrm{67.9keV}$ $\mathrm{^{44}Ti}$ feature at 2300 days (the end of the optical observations in 2017, as presented here) as $1.6 \times 10^{-5} \,\mathrm{photons}\,\mathrm{cm^{-2}s^{-1}}$.  Despite the distance to M51, the young age of SN~2011dh and the significant quantity of $\mathrm{^{44}Ti}$ implied would make this feature detectable in modest NuSTAR observations \citep{2015ApJ...814..132L}.

\subsection{Origin of the light echo scattering medium}
\begin{figure*}
\begin{center}
\includegraphics[width=15cm]{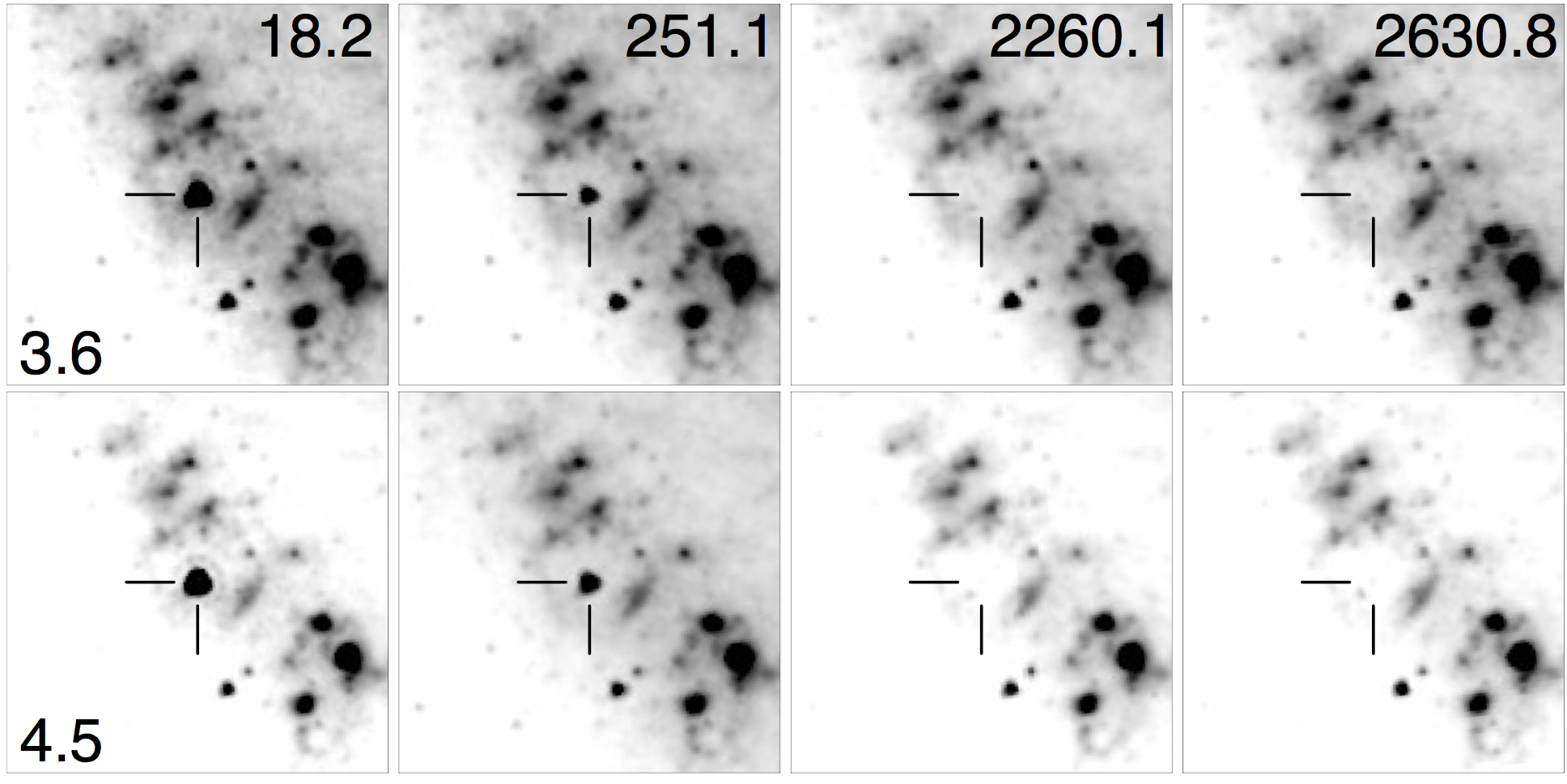}
\end{center}
\caption{{\it Spitzer} observations of the site of SN~2011dh at early and late-times, in the $3.6\mu m$ ({\it Top Row}) and $4.5\mu m$ ({\it Bottom Row}) bands.  Each panel is centred on the position of SN~2011dh, oriented such that North is up and East is left and has dimensions $80 \times 80\arcsec$.}
\label{fig:disc:spitzer}
\end{figure*}

Previously, \citet{2013ApJ...778L..19H} reported the observation of a ``thermal" echo arising from SN~2011dh, observable as light from heated grains (from the shock breakout and the SN itself) emitted at mid-infrared wavelengths up to 625 days post-explosion.  This dust extended in distance from $\sim 0.1\mathrm{pc}$ to $\sim 0.26\mathrm{pc}$ from the SN.  We examined more recent {\it Spitzer} Infrared Array Camera observations of M51, covering $1029 - 2630\mathrm{d}$ post-explosion available in the {\it Spitzer} Heritage Archive\footnote{https://sha.ipac.caltech.edu/applications/Spitzer/SHA/}.  We followed the photometric procedure of \citeauthor{2013ApJ...778L..19H}, using zeropoints presented in the {\it Spitzer} instrument handbook\footnote{https://irsa.ipac.caltech.edu/data/SPITZER/docs/irac/iracinstrumenthandbook/home/} and aperture corrections derived from the observations themselves.  After the period of observations presented by \citeauthor{2013ApJ...778L..19H}, we do not significantly recover a single source, at epochs overlapping our HST observations, at the SN position at either $3.6 \mu m$ or $4.5 \mu m$ (see Figure \ref{fig:disc:spitzer}) after the detections reported by \citeauthor{2013ApJ...778L..19H}.  

The dust detected, through the thermal echo, at early times is different in spatial distribution compared to the geometric distribution required to produce an optical light echo at late-times (at which time there is no significant signal at mid-infrared wavelengths in the late-time {\it Spitzer} observations).  As \citeauthor{2013ApJ...778L..19H} notes, the early time thermal echo must arise from pre-existing dust and the distance scales required of the late-time echo similarly require pre-existing dust.  That the progenitor of SN~2011dh should be surrounded by significant quantities of pre-existing dust over a variety of different spatial scales hints at the complex mass loss history experienced by the star.

Our preferred geometric model for the production of the late-time brightness of SN~2011dh through a light echo is for the dust to be arranged in an inclined disk configuration.  For a disk with latitudinal opening angle $\Theta_{d}$, the corresponding solid angle subtended by the disk is $\Delta \Omega = 2\pi \times 2 \times \cos\left(\pi/2 - \Theta_{d}\right)$.  The total hydrogen mass contained in the disk is, therefore, given by:
\begin{equation}
M_{tot} = \int^{R_{2}}_{R_{1}} 4 \pi r^{\prime 2} \left[ n_{0}(H)\left( \frac{r^{\prime}}{R_{1}}\right)^{- \alpha}\right] \mathrm{d}r^{\prime} . \frac{m_{H}}{M_{\odot}} . \frac{\Delta \Omega}{4 \pi}
\end{equation}
where $m_{H}$ is the mass of a hydrogen atom.

Given the high-densities ($n_{0}(H) \sim 250 - 275\,\mathrm{cm^{-3}}$) and the large spatial extent ($R_{1} \sim 2\,\mathrm{pc}$) we find that the scattering medium would correspond to total hydrogen mass of $M_{H} \gtrsim 35M_{\odot}$ (for constant density).  Given the initial and final masses of the progenitor \citep[$\sim 13 M_{\odot}$ and $3 - 4M_{\odot}$, respectively;][]{2011ApJ...739L..37M,2012ApJ...757...31B}, if the scattering medium were composed of material just lost from the progenitor then the dust-to-gas mass ratio would need to be larger by a factor of $\gtrsim 4 - 5$ times higher than we have assumed \citep[adopting the ratio of dust-to-gas consistent with the interstellar average of $\sim 0.01$;][]{2007ApJ...663..866D}.  This value is similar to that seen in the winds of red supergiants \citep[$\sim 0.005$;][]{2016MNRAS.462.2995M}, which suggests that such an enhancement of the dust-to-gas mass ratio, required if the material originated from the progenitor, is unlikely.

The mass, spatial extent and the densities inferred are consistent with examples of wind-blown bubbles \citep{2015A&A...584A..49V}.  Given a wind velocity of $\sim 20\,\mathrm{km\,s^{-1}}$ appropriate for YSG mass loss and as measured for SN~2011dh \citep{2014ApJ...785...95M}, such a structure could have been created within $\sim 50 - 100 \times 10^{3}\,\mathrm{yr}$ prior to explosion, which is consistent with the timescales of the enhanced mass loss predicted for the primary star in the binary progenitor system immediately before the SN \citep{2013ApJ...762...74B}.  The medium around SN~2011dh could be considered to be composed of two phases: i) a nearby, smooth CSM (with density $\propto r^{-2}$) \citep{2014ApJ...785...95M,2016MNRAS.455..511D} composed of freely expanding mass lost from the progenitor prior to explosion; and ii) and a denser, shocked interface between wind material and the surrounding interstellar medium at large radii \citep{2005ApJ...630..892D}.  Such structures, and their interaction with the ejecta, are expected to drive the shapes of SN remnants as observed 100s of years after explosion \citep[see e.g. Cygnus Loop;][]{1997ApJ...484..304L}.  Given the current free expansion of the SN \citep[$\sim 19\,000\,\mathrm{km\,s^{-1}}$; measured in the radio using the Very Large Array by ][]{2016MNRAS.455..511D}, the ejecta may eventually interact with the material responsible for the light echo within $\sim 100\,\mathrm{yr}$ and, as such, the late-time behaviour of light curve we have observed may serve as a precursor to the emergence of a SN remnant.  If this structure is a wind-blown bubble, then its non-spherical shape has important implications for the geometry of mass loss from the progenitor or the role of magnetic fields in the interstellar medium into which it is expanding \citep{2015MNRAS.453.4467M,2015A&A...584A..49V}.

As the preferred models locate the dust almost in the plane of the sky, and not along the line of sight, the presence of such dust has no implications for the reddening inferred towards the SN or the progenitor themselves.  The existence of such a localised, dense structure could contribute to the differential extinction that is often observed for massive star populations in other galaxies \citep{2018MNRAS.476.2629M}; even for the most extreme disk model ($n_{0}(H) = 275\,\mathrm{cm^{-3}}$, $R_{2} = 2.7\,\mathrm{pc}$ and $\Theta_{d} = 17.5^{\circ}$) the additional reddening that would be induced for a background star across the thickest part of the disk would only be $E(B-V) = 0.3\,\mathrm{mag}$.  The edges of these bubbles, around massive stars, could be important yet invisible components of the resolved stellar populations observed around the SNe resulting from the explosions of massive stars.

\subsection{Implications for the binary companion}
\begin{figure}
\begin{center}
\includegraphics[width = 8cm]{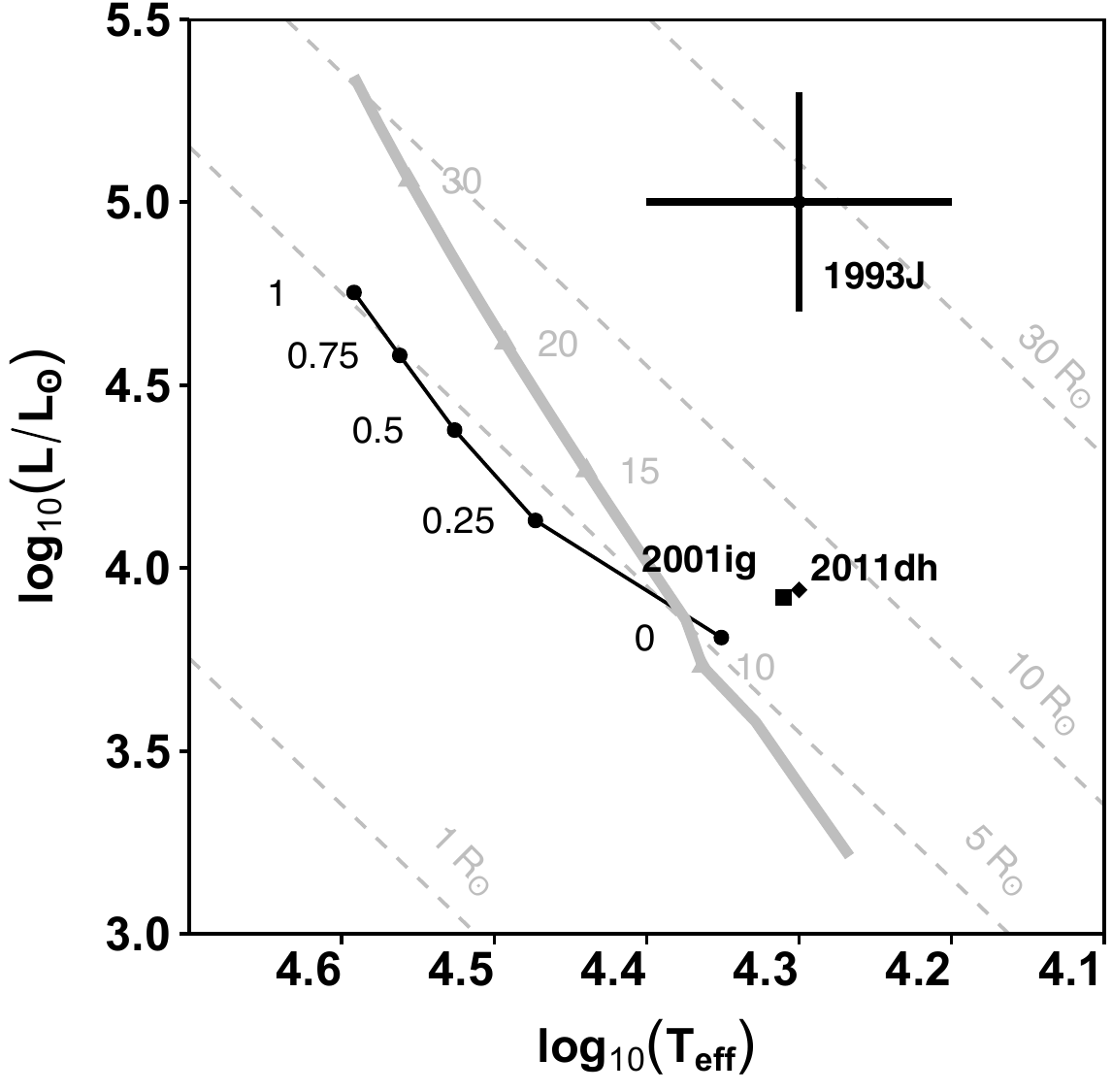}
\end{center}
\caption{Hertzprung-Russell diagram showing the locations of companion stars to the progenitors of the Type IIb SNe 1993J \citep{maund93j}, 2001ig \citep[$\blacksquare$][]{2018ApJ...856...83R} and 2011dh ($\blacklozenge$; presented here).  Over-plotted, in black, are the models of the companion star for different values ($\beta$) of the fraction of mass lost from the progenitor that was accreted onto the companion \citep{2013ApJ...762...74B}, which for these observations implies $\beta = 0$.  The dashed grey lines indicate lines of constant stellar radius in units of solar radii, as indicated on the figure.  The solid grey line indicates the main sequence for single stars \citep{2017PASA...34...58E} from which the companions, as B-supergiants, are clearly offset.}
\label{fig:hrd}
\end{figure}

For the calculation of the properties of the binary companion on the Hertzprung-Russell diagram (see Figure \ref{fig:hrd}), we considered the photometry of the companion, derived from the flux difference at UV wavelengths between the late-time observations and the light echo models, with respect to the ATLAS9 library of stellar spectra appropriate for supergiant stars \citep{2004astro.ph..5087C}.
Given the reddening inferred towards SN~2011dh, the flux deficits observed in the $F225W$, $F336W$ and $F435W$ filters are consistent with a blue supergiant star with $\log (L/L_{\odot}) = 3.94 \pm 0.13$ and $\log (T_{eff})= 4.30 \pm 0.06$ (see Fig. \ref{fig:hrd}), corresponding to a mass of $\sim 9 - 10M_{\odot}$ \citep{2017PASA...34...58E}.  This is consistent with the prediction of \citet{2013ApJ...762...74B} for a system in which the companion does not accrete material from the progenitor.  As discussed by \citet{2015MNRAS.454.2580M}, the interpretation of the 2014 UV observations, in which \citet{2014ApJ...793L..22F} claim a detection of the companion, need to be modified to include a non-negligible contribution from the lightecho.   While this star is significantly fainter than the companion identified for SN 1993J \citep[initially $\sim 14M_{\odot}$, but reaching $22M_{\odot}$ by the time of explosion;][]{maund93j}, it is coincidentally almost identical to the companion inferred for SN~2001ig \citep{2018ApJ...856...83R}.  Unlike the companion of SN~1993J, the companion to SN~2011dh may have induced additional mass loss from the progenitor but did not accrete that material itself (see Fig. \ref{fig:hrd}).   The identification of another companion within a restricted temperature range, appropriate for B-supergiants, could mean that a survey for B-supergiants in close proximity to nearby resolved SN remnants could potentially reveal companions to SNe whose progenitors evolved in binary systems.

\section{Conclusions}
We have presented late-time HST photometry of the Type IIb SN 2011dh, extending to $2299\mathrm{d}$ post-explosion.  After the initial decline at early times, the light curve exhibited a plateau and then after $t > 1950\mathrm{d}$ exhibited a further decline.  The rate of the decline of the bolometric light curve is much shallower than would be expected if the SN luminosity was just powered by the decay of $\mathrm{^{56}Ni}$ and its daughter products. Narrow-band imaging of the SN at the wavelengths corresponding to $\mathrm{[ O\, I ]\,\lambda 6300}$ and $\mathrm{H\alpha}$ do not show evidence of the emergence of strong emission lines, implying that interaction of the ejecta with a dense CSM is unlikely to be the cause of the late-time luminosity.  The flatness of the bolometric light curve and its subsequent decline are shallower than the predictions for a magnetar powered light curve.  The behaviour of the light curve can be described by the decay of radioactive elements, in particular a mixture of $\mathrm{^{56}Co}$ and $\mathrm{^{44}Ti}$, but not a significant quantity of $\mathrm{^{57}Ni}$; however the quantity of titanium required would necessitate SN~2011dh producing more in the explosion than any previous Type IIb SN, by at least an order of magnitude.  A late-time light echo, in a disk-like configuration almost in the plane of the sky, can explain the overall brightness evolution of the light echo, including the initial plateau (and possible re-brightening), and the later decline, as well as the evolution of the colour.  The dust required to produce this light echo would have to be located over spatial scales of order $\sim 1.8\,\mathrm{pc}$ to $\sim 2.7\,\mathrm{pc}$, which could correspond to a wind-blown bubble formed around the progenitor prior to explosion.  Assuming a light echo as the origin of the late-time brightness of the SN, we extracted the photometric properties of the companion and constrained it to be a $9 - 10M_{\odot}$ star.

\acknowledgments
The research of JRM is supported through a Royal Society University Research Fellowship.  JRM is grateful to Yi Yang for helpful discussions during the preparation of this manuscript.
Based on observations made with the NASA/ESA Hubble Space Telescope, obtained from the data archive at the Space Telescope Science Institute. STScI is operated by the Association of Universities for Research in Astronomy, Inc. under NASA contract NAS 5-26555.  HST data were obtained 

\software{DOLPHOT \citep{2016ascl.soft08013D}, PyRAF \citep{2012ascl.soft07011S}}

\bibliographystyle{aasjournal}

\end{document}